\documentclass[sigconf,nonacm]{acmart}
\usepackage{subcaption}
\usepackage[inline]{enumitem}
\AtBeginDocument{%
  }

\setcopyright{none}
\begin{document}

\title{Progressive Web Application for Storytelling Therapy Support}


\author{Javier Jiménez-Honrado}
\email{javier.jimenez19@alu.uclm.es}
\author{Felipe Costa-Tebar}
\email{felipe.costa@uclm.es}
\affiliation{%
\institution{I3A (UCLM)}
\city{Albacete}
\country{España}}

\author{Javier Gómez}
\email{javier.gomez21@alu.uclm.es}
\author{Felix A. Marco}
\email{felix.albertos@uclm.es}
\affiliation{%
\institution{FCSTI (UCLM)}
\city{Talavera}
\country{España}
}

\author{Jose A. Gallud}
\email{jose.gallud@uclm.es}
\author{Gabriel Sebastián Rivera}
\email{gabriel.sebastian@uclm.es}
\affiliation{%
\institution{ESII (UCLM)}
\city{Albacete}
\country{España}}


\renewcommand{\shortauthors}{JImenez-Honrado et al.}

\begin{abstract}
In spite of all advances promoted by information technologies, there are still activities where this technology is not applied for reasons such as being carried out in non-profit organizations or because they have not adapted to this modernization. Until recently, the way to work with mobile devices was either by connecting through a web page with the device's browser, or by downloading an application from the corresponding platform. But lately, technologies are being developed that aim to break with this, as in the case of Progressive Web Applications (PWA). One of the advantages offered by PWA is to access the web page and install it as an application on the device. The purpose of this article is to design a progressive Web application for the support of Storytelling Therapy, one of the novel therapies applied in the field of mental health. In addition to providing a software application to enhance Storytelling Therapy workshops, it is also intended to analyze and verify the advantages of PWA in a real case. 
\end{abstract}

\ccsdesc[500]{Human-centered computing~Human computer interaction (HCI)}
\ccsdesc[500]{Human-centered computing~Accessibility}
\ccsdesc[500]{Human-centered computing~Interactive systems and tools}

\keywords{Progressive Web Application, Interactive Systems Design, Storytelling Therapy, Storytelling in education}

\received{20 February 2007}
\received[revised]{12 March 2009}
\received[accepted]{5 June 2009}

\maketitle

\section{Introduction} \label{introduction}
Storytelling therapy with children is a therapeutic approach that harnesses the power of narratives to facilitate emotional expression, promote self-discovery, and foster resilience in young minds \cite{Marleen2008}. Through the art of storytelling, children can explore and communicate their thoughts, feelings, and experiences in a safe and imaginative space. This therapeutic method not only helps children make sense of challenging situations but also enhances their cognitive and emotional development \cite{Ochs1992} \cite{Isbell2004} \cite{Skinner2007}. By engaging in the creation and sharing of stories, children gain valuable insights into their own strengths and coping mechanisms. This approach proves to be especially beneficial for children facing various emotional and behavioural challenges, as it provides them with a creative outlet for self-expression, encourages problem-solving skills, and ultimately contributes to their overall mental well-being.

This article presents an innovative application that seeks to merge therapy through popular stories and myths with the emerging technologies provided by the irruption of Progressive Web Applications \cite{pwa} (PWA). The proposed system allows users to create stories from a set of characters and events, according to their own criteria. Once created, the story is summarized and the user is guided by a therapist in the process of reflection and drawing conclusions. This approach assists one of the techniques used in a local association “\textit{Escuela de Cuentoterapia de Albacete}”, integrating it into a solution that leverages the features of PWAs. Our proposal represents a significant step in combining traditional therapies with emerging digital technologies, providing a valuable tool for the association, which is dedicated to conducting therapies, mainly with children, using popular stories and myths.

The application has been designed following the guidelines indicated by experts from the association and the final prototype has been evaluated with a group of participants, some of them experts in Storytelling Therapy.

The article is organized as follows. The background of this work is divided into two sections. Section \ref{background} presents basic concepts about Storytelling Therapy and Section \ref{pwa} is devoted to explaining the Progressive Web Application technology. Section \ref{swarch} describes the software architecture designed for this prototype. Section \ref{app} describes the CuentoterApp application, including the main functionalities. The usability evaluation is presented in section \ref{evaluation}. Finally, section \ref{conclusions} summarizes the conclusions and future work.

\section{Background and basic concepts} \label{background}
Vladimir Yákovlevich Propp, Russian anthropologist and linguist; and Joseph John Campbell, American mythologist, writer and professor, carried out an analysis of the stories and myths (respectively) from the entire history of humanity and many different cultures \cite{propp1998morphology} \cite{campbell2020heroe}. After this analysis, they concluded that there is a set of fixed patterns that all of them meet so that we have a set of "actants" (characters) and a set of "functions", which are the different events that can happen in the story, as:

\begin{itemize}
    \item “Estrangement from the family”: the hero of the story separates from his family for some reason.
    \item “Deception”: the antagonist manages to deceive the protagonist for his benefit.
\end{itemize}
\vspace{-1mm}
From these functions, a pattern of similarity emerges across 31 distinct functions. While not all functions are present in every story or myth, a subset invariably exists and maintains a consistent order within the set of 31. 

Starting from these concepts, Giovanni “Gianni” Rodari, an Italian educational writer and journalist (specialized in children's and youth literature), devised a card game based on the creation of stories using the concepts of functions and actants \cite{rodari2002gramatica}.

“Storytherapy” is based on doing different dynamics and workshops related to stories, each one with different characteristics and objectives, but with therapeutic purposes \cite{storytherapybenefits}. For this research, the aim is to emulate and work on one in particular.

"Storytelling therapy" has been used as a resource aimed at children between 0 and 6 years old in cases such as hospitalization due to illness to alleviate their pedagogical maladjustment \cite{cabrera2022cuentoterapia}. An example where "storytelling therapy" has been used is in the study carried out by Mohammed et al., this study consisted of the treatment of cases of anxiety and depression in 137 patients between 9 and 13 years old who had cancer, giving results that in the context where the study was carried out, it was a success to use storytelling therapy, although it should be mentioned that they also used writing therapy \cite{Mohammed2022}.

This article, mainly applicable to children aged 0-6 and 9-13, emphasizes the method for those aged 6+. It uses Rodari’s game, where participants choose characters and craft narratives using cards. The therapist aids in this process. This aligns with Mohammed et al.'s findings on the therapeutic benefits of story writing, which builds on prior research showing anxiety reduction through creative practices like drawing and writing \cite{Altay2017}.

The objective of the dynamic is not to give a moral or explanation to the child, but rather to assist him in reflecting on the ideas that he has chosen in these different situations; but always with the focus of drawing his reflections.

In the context of this research study, it is pertinent to acknowledge the representation of Storytherapy in Spain, embodied by the ‘Asociación Iberoamericana de Cuentoterapia’. This association, established in 2010, is dedicated to the promotion of Storytelling and its pedagogical principles, in addition to accrediting teachers and institutions in the field of storytelling therapy.

We are privileged to have established a connection with an expert affiliated with the ‘Escuela de Cuentoterapia’ in Albacete. This association is also engaged in conducting workshops centered around the dynamics of Storytelling Therapy. The expert in question has been instrumental in shaping the proposal and has provided invaluable information that forms the foundation of this study.

In this context, attention is also given to its digital representation through various applications that embody the principles discussed, providing insights into its potential in the digital realm. Such is the example of Storyjumper, a web application, facilitates story creation and creativity, offering publishing capabilities through a user-generated library. Central to our discussion is its story development editor, an interactive tool supporting image, text, and sound integration on each page \cite{Bagua2023} \cite{fitriyani2023}. It features a template character creator and allows for both application-provided and personal images. Finished works are downloadable or re-editable. 

Another group of researchers, Garzotto et al., devised a comparable setting known as FaTe2, an edutainment web based environment for children that incorporates storytelling, hypertext, games, collaborative learning, and social interaction \cite{Garzotto2006}.

Carter et al.'s findings indicate that using these templates for character creation shows no age-related style preferences \cite{Carter2016}. However, a limitation is the incompatibility with Android devices, restricting use to iPads and computers. Furthermore, the editor's design and layout are not optimized for smaller screens, such as smartphones.

BookTraps \cite{booktraps} is a children’s app for creating mini-books primarily through drawings, available only on Android and iOS, not as a browser application. Its user interface is specifically tailored for mobile devices. PicaBook \cite{picabook} also a mobile-only app, offers a broader story creation platform, supporting text, images, stickers, sound, and a publishing library in addition to drawing.

This review underscores the absence of an application comparable to our proposed one, especially one leveraging PWA (Progressive Web App) technology.

\section{Progressive Web Applications} \label{pwa}
Traditionally, mobile apps were written in Android and iOS native languages, requiring more effort but fully utilizing device capabilities. Mobile web development, while similar to regular web page creation, lacked the full OS access of native apps.

Over the years, different approaches have been addressed to try to unify the development effort for all platforms, but without losing the capabilities of each device. For example: (a) Electron.js: an open-source framework for creating desktop applications with code in JavaScript, HTML and CSS; (b) Apache Cordova or Capacitor: frameworks that compile code in JavaScript, HTML and CSS, and wrap it in native containers for each platform, so that it is written with those languages and generates the corresponding native application; and (c) Progressive Web App: a philosophy and set of technologies to unify native and web.

PWA, are a type of application that aims to be a middle ground between web and native apps. Native apps have the advantage of taking better advantage of the device's features, making better use of the operating system, being able to work with them offline \cite{fam2017jwe} or being able to access features such as the USB port or a camera. PWA also made possible the unification for the end-user app experience, including, but not limited to, installable and native-looking web apps \cite{biarnhansen2017}, which make them an affordable alternative to a native application that may be considered \cite{ming2022}. Also, as mobile devices are limited concerning battery capacity, developers should keep the energy footprint of a mobile app as low as possible. As Huber et. al. pointed out \cite{huber2022}, PWAs can be considered a viable alternative in terms of energy efficiency to other mobile cross-platform development approaches.

A PWA’s progressiveness is determined by its fulfilled characteristics: Detectability, Installability, Linkability, Progressive Enhancement Support, Network Independence, Responsive Design, and Security. Next, essential features like Service Worker, Installability, and testing tools like Lighthouse are explained.


\textbf{Service Workers}

To support the features of being able to work offline or push notifications, service workers are used. These are virtual proxies that act as an intermediary between the browser and the network, intercepting requests and deciding whether to handle them from some local algorithm or from the network \cite{serviceworkers} \cite{pwaoffline_2023}.

They operate on a different thread than the program and must be registered first, and they only act on the resources that you specify them to act on. In this way the program continues to run or at least part of its functions, when there is no internet connection or outages occur. Not only do they allow offline, but also features such as push notifications or perform very expensive algorithms in a different thread so as not to compromise the normal execution thread. However, being so powerful, they have a disadvantage and that is that their security cannot be compromised, which is why they only work in a secure connection context such as HTTPS \cite{pwaoffline_2023}.\newline

\textbf{Installation capability}

PWAs possess the capability to install themselves, a feature that significantly enhances the user experience. For the user, the web application ceases to be a tab in the browser and becomes a normal application on the home screen of the mobile as he is accustomed to. In order to use this feature it will be necessary to have service workers and a secure connection as already explained, and it is also necessary to elaborate the Web Manifest. The Web Manifest is a file in ".json" format that lists all the information of the application. Information such as the title, icons of different sizes, background colour for the loading and startup screens... \cite{pwasinstallable_2023}.

Upon meeting all requirements and with browser support, an installation option for the application appears when the browser configuration button is clicked. Post-installation, the website presents as a standard app without a URL.\newline




\textbf{Lighthouse}

"Lighthouse" is a tool of the Chrome development tools. This can be run from the browser on any website and serves to improve the quality of web applications. The tool is responsible for passing a series of tests and generating a report with what meets and fails on the web and how to improve it. Among these tests is one that measures what percentage of the application is progressive and what are the features with which it complies. It also justifies why these tests should be there and the advantages they provide. 


\section{Software Architecture} \label{swarch}
It was decided to implement a solution with a Single-Page Application (SPA) model, but with the particularities of PWA, which has been previously used in mobile and e-learning scenarios \cite{fam2015hci}. A SPA application is defined by having the control logic on the client side and communicating with the server at the beginning so that it loads in part and from the client it is processed and displayed as needed.

You can also have asynchronous communications from time to time to retrieve some data in JSON format and update reactively. This approach is advantageous since it provides a better user experience since it does not have to be reloaded with every action and also loads faster. 

\begin{figure}[H]
  \includegraphics[width=0.6\linewidth]{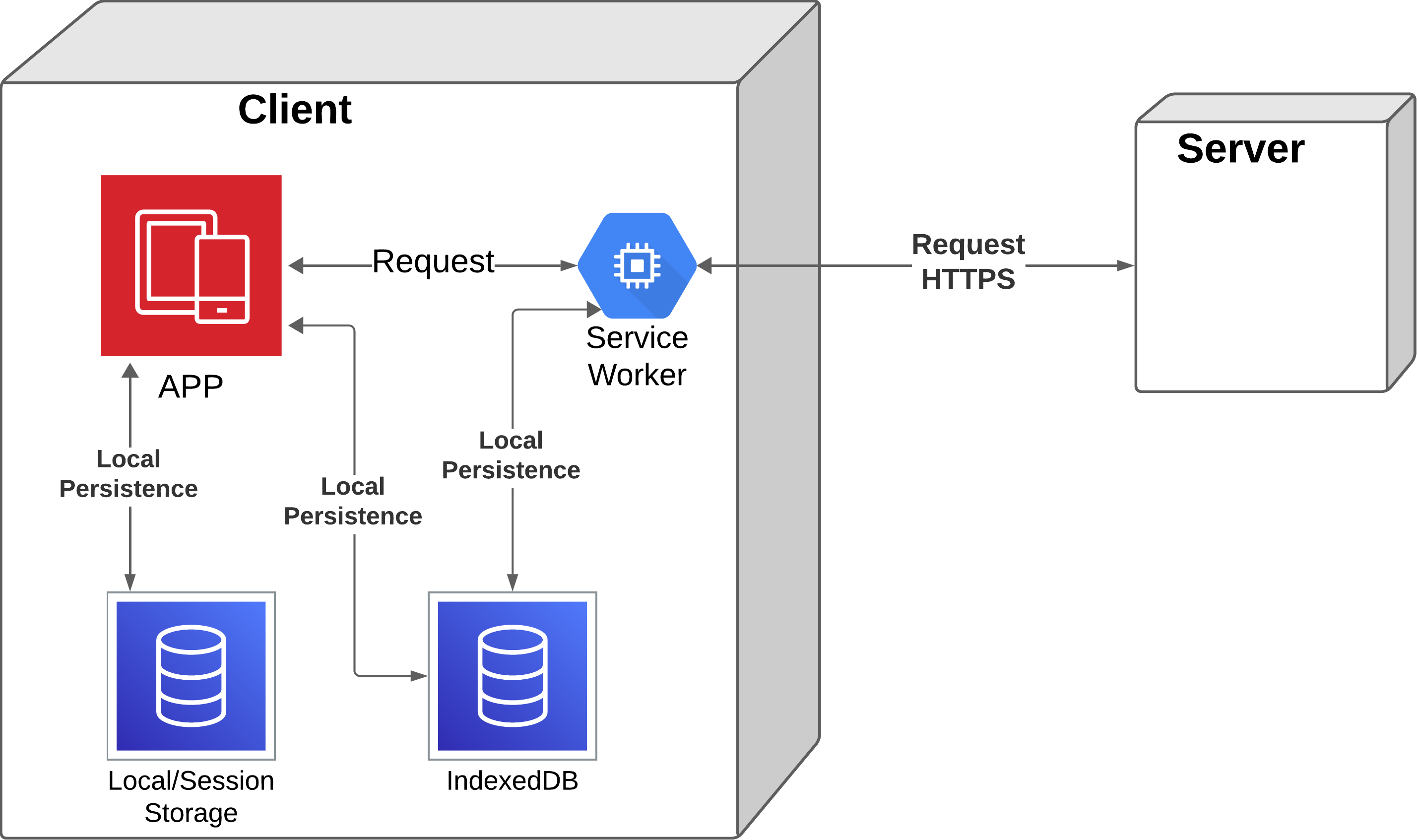}
  \caption{Software architecture}
  \Description{Software architecture}
  \label{fig:swArch}
\end{figure}

In the case of PWA, it works in a very similar way, except that in this case, we will also have a service worker (which acts as a virtual proxy) in the middle, which intercepts and manages the client's requests, so that it allows features such as offline use or installation. The offline case allows it since it has a cache and internal browser databases. A diagram of the structure to be used in the proposal is shown in Figure \ref{fig:swArch}.

Generally, the client will communicate with the server as a last option, before checking if the necessary information is in local. To carry out this structure we will use Angular \cite{angular}, a framework and development platform for the creation of SPAs and Ionic, a software development kit for the creation of hybrid applications \cite{ionic}, which has libraries to work with PWA. Next, it will be shown which PWA features are interesting for the application.

The next step is to analyze and select the desirable features of PWA to assist in Storytelling Therapy.

Initially the use of the application will be in workshops or therapy dynamics and will be carried out in spaces such as schools, libraries or association halls, not knowing the available material or internet connection in these places. Therefore, each PWA characteristic should be analyzed to decide whether or not is included in our prototype:
\begin{itemize}
    \item Detectable: this is a desirable feature as it facilitates access to the application and aids its dissemination, but it is not entirely necessary operation.
    \item Installable: this feature is absolutely necessary as it allows us to leverage all the features of PWA, enhances the user experience, and simplifies its use for individuals such as children.
    \item Linkable: this feature is also completely necessary so that in the context of a workshop the system can be easily shared.
    \item Progressive enhancement support: mandatory feature as we do not know what devices are available in these dynamics and workshops. Therefore, it must work on the largest number of available devices, regardless of their capabilities.
    \item Re-engageable: this feature is not desirable as the context of use will typically be in the workshop. Consequently, there is no need to be notified to re-enter the application outside this context. However, if the project’s needs scale, it could become interesting.
    \item Network independent: this feature is also completely mandatory as we cannot guarantee that there will be a Wi-Fi connection or even coverage in the spaces used. Therefore, the functionalities, or at least part of them, must work in a context without an internet connection.
    \item "Responsive" design: like progressive enhancement support this is also necessary since it is necessary that the system works regardless of the device resolution. However, it is arguable that it currently makes no sense to create a web application without a responsive design, as most of the devices used today are mobile.
    \item Secure: with what is currently proposed, it does not seem a desirable feature since sensitive information will not be handled. However, if the project’s needs scale, it may become necessary.
\end{itemize}

After this list of desirable features, we can argue that the use of PWA is very convenient for a project with these characteristics, besides the features defined as desirable will be used as a measure of the quality of the project and as a requirement that it is "done".

\begin{figure}[H]
  \includegraphics[width=0.75\linewidth]{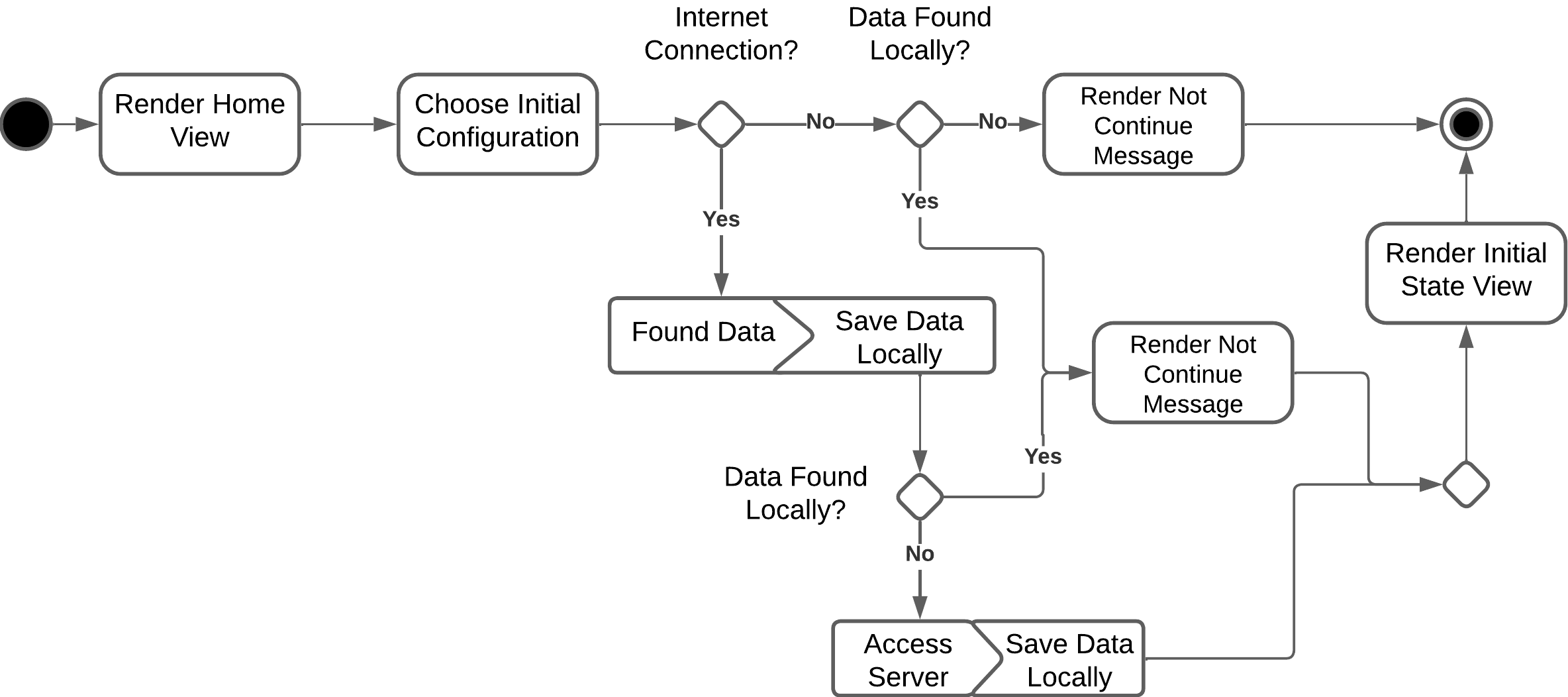}
  \caption{Activity diagram}
  \Description{Activity diagram}
  \label{fig:activity}
\end{figure}

The final development, as already mentioned in the previous chapter, consists of a SPA solution together with the explained particularities offered by PWA, so that an initial request is made to the server through the "service worker" and from there either the local storage is consulted or requests are made to the server, depending on whether the information exists or not (locally). A more visual example of this situation can be seen in the activity diagram in Figure \ref{fig:activity}.

In this case, we can check how the internal logic works depending on whether the data is available or not for the case of choosing the initial situation for the story. Thus, if the data is available and there is no internet connection, the user should not notice it and everything should work normally. In the example above (see Figure \ref{fig:activity}) a specific component is shown, but for the rest it works in a very similar way.

Local persistence is developed using "IndexedDb". This is a NoSQL database hosted in the browser and whose storage depends on the capacity of the device as well as the browser.

In this development the logic part has been developed in the client, this is possible thanks to using the "Angular 15" framework, which allows a component-oriented development, where in these components the view parts and the logic that manages them can be hosted. In addition, another framework that works on Angular is used for the view, this is "Ionic 6", which provides components that facilitate the development of a more compatible interface for mobile devices.

\begin{figure}[H]
  \includegraphics[width=0.75\linewidth]{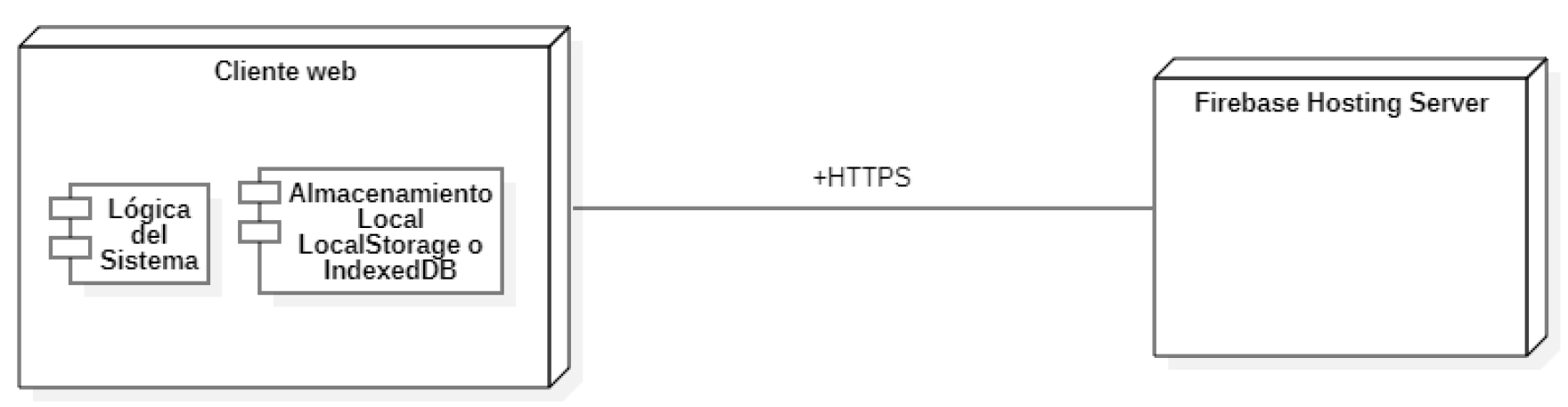}
  \caption{Deployment diagram}
  \Description{Deployment diagram}
  \label{fig:deploy}
\end{figure}

Finally, a library, "Dexie.js" \cite{dexiejs}, is used to facilitate the use of IndexedDb and the API "Web Speech API" \cite{web_2023}, which provides an interface to recognize and translate from voice to text the microphone on the web.
The deployment is done through "Firebase", a platform for free web hosting that belongs to Google, thanks to this you get hosting with HTTPS protocol, in the deployment diagram in figure \ref{fig:deploy} can be seen in a more representative way.

It should be noted that the web browser can be on different platforms such as Windows, Mac OS, Android or iOS.

\section{The CuentoterApp Application} \label{app}

CuentoterApp is a solution that allows the creation of stories imitating the procedure of one of the therapies of the 'Escuela de Cuentoterapia' of Albacete, from an initial situation and some chosen characters, you can go writing the story fragment by fragment, where each fragment has a thematic framework.

Once the story is finished, the user has the option of downloading it in PDF format or saving it in a "Library", a local storage of all the stories written by the author.

The solution is intended to support this therapy process so that it replaces the traditional way in which the story is recorded. It is intended to strengthen the user's creativity, highlight the importance of stories, show the virtues of a PWA approach and take advantage of its benefits.

In this section we explain how the system works from the user's point of view, explaining it step by step with illustrative images.

\begin{figure}[H]
	\centering
	\begin{subfigure}{0.3\linewidth}
        \centering
		\fbox{\includegraphics[width=0.6\textwidth]{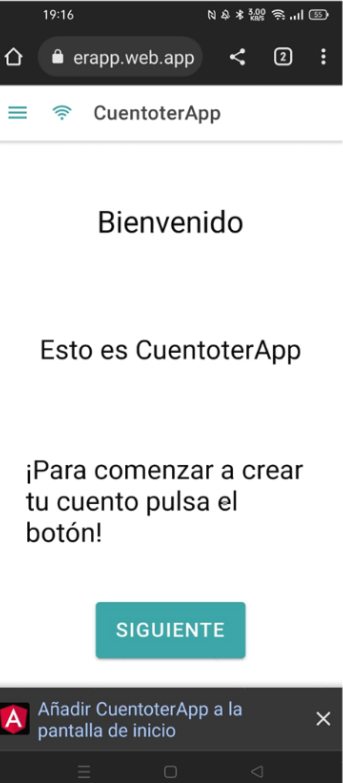}}
		\caption{Welcome Page}
		\label{fig:welcomPage}
	\end{subfigure}
	\begin{subfigure}{0.3\linewidth}
        \centering
		\fbox{\includegraphics[width=0.6\textwidth]{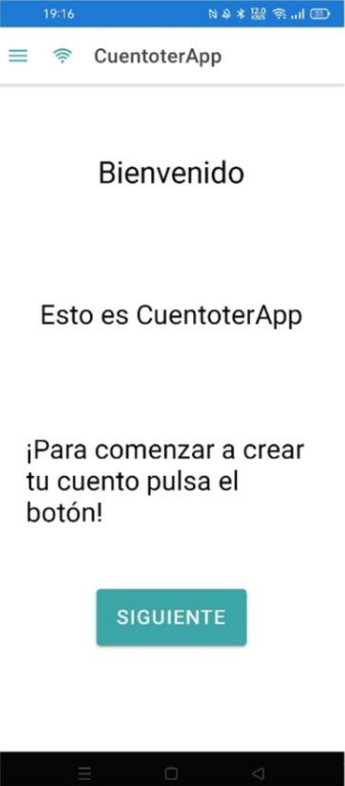}}
		\caption{Add To Home}
		\label{fig:addToHome}
	\end{subfigure}
    \begin{subfigure}{0.3\linewidth}
        \centering
		\fbox{\includegraphics[width=0.6\textwidth]{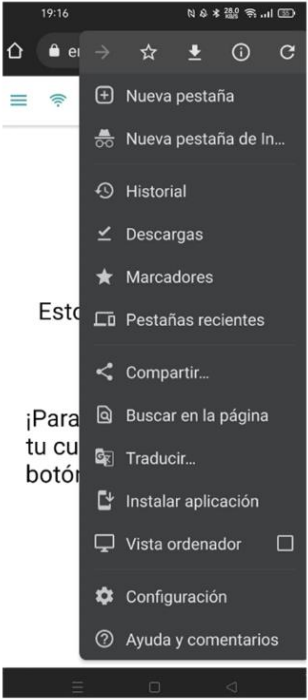}}
		\caption{Install App}
		\label{fig:instApp}
	\end{subfigure}
	\caption{App Description}
	\label{fig:CuentoterApp}
\end{figure}

The first thing you see when entering the address mentioned in the previous section is a welcome page (Figure \ref{fig:welcomPage}). At this point there are two possibilities, install the application or use it directly from the browser. Suppose it is the case of installing it. In that case, either a browser message may appear indicating if we want to add the web to the home page (Figure \ref{fig:addToHome}) or it may not appear. We must go directly to the browser options and if it is compatible, an option to "install the application" should appear here (Figure \ref{fig:instApp}).

\begin{figure}[H]
	\centering
	\begin{subfigure}{0.3\linewidth}
        \centering
		\fbox{\includegraphics[width=0.6\textwidth]{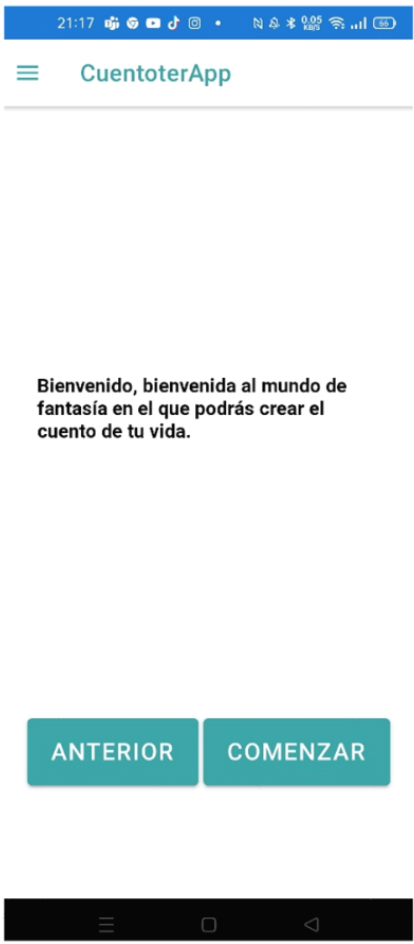}}
		\caption{Create Story}
		\label{fig:startPage}
	\end{subfigure}
	\begin{subfigure}{0.3\linewidth}
        \centering
		\fbox{\includegraphics[width=0.6\textwidth]{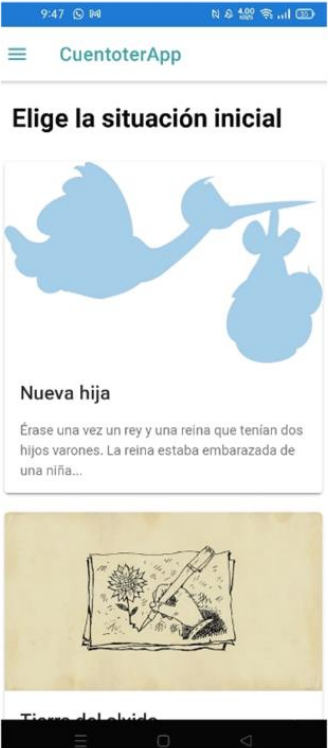}}
		\caption{Initial Situation}
		\label{fig:firstSituation}
	\end{subfigure}
    \begin{subfigure}{0.3\linewidth}
        \centering
		\fbox{\includegraphics[width=0.6\textwidth]{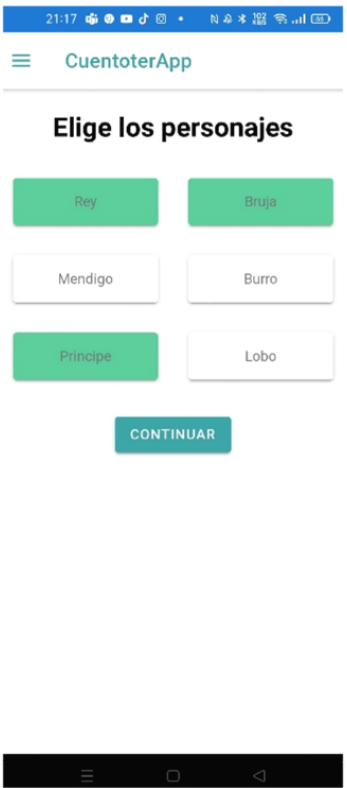}}
		\caption{Character Choice}
		\label{fig:characterChoice}
	\end{subfigure}
	\caption{Application Flow}
	\label{fig:appFlow}
\end{figure}

After this to follow the basic flow of the system (create story), it will be necessary to click on the "NEXT" button, this will lead to a series of pages with a welcome and explanatory text. It is necessary to click the button until the last one is reached (Figure \ref{fig:startPage}).

If you click on the "PREVIOUS" button, you will go back in these pages and to start the process of creating the story, click on the "START" button. When you click on it, you will go to the page for choosing the initial situation (Figure \ref{fig:firstSituation}), here several cards appear with a photo, title and description, and by clicking on them you can choose the one you want as the context for the story.

Next, the character selection screen will appear (Figure \ref{fig:characterChoice}), click on all the character names you want to add (the selected ones are marked in green), and then press the "CONTINUE" button.


\begin{figure}[H]
	\centering
	\begin{subfigure}{0.3\linewidth}
        \centering
		\fbox{\includegraphics[width=0.6\textwidth]{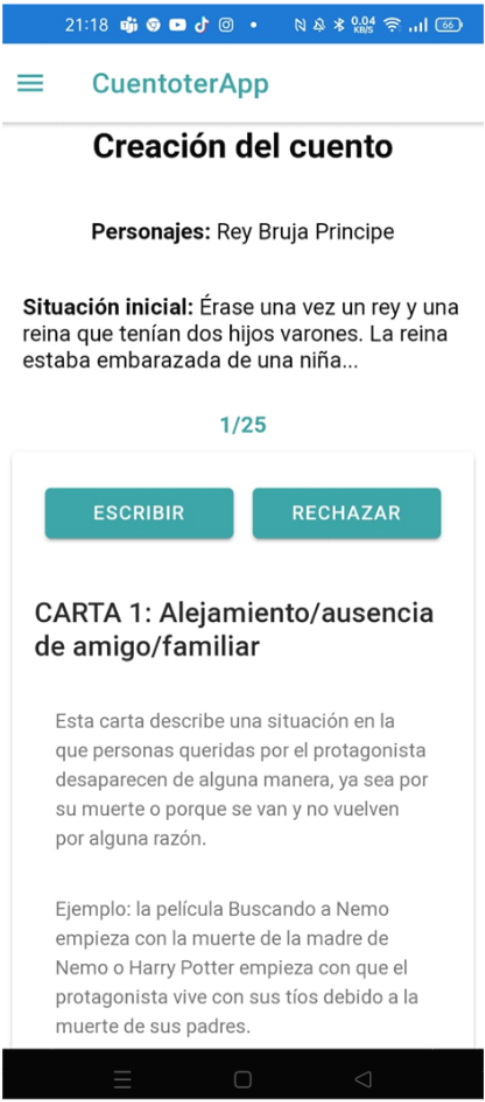}}
		\caption{Description}
		\label{fig:descriptionFunction}
	\end{subfigure}
	\begin{subfigure}{0.3\linewidth}
        \centering
		\fbox{\includegraphics[width=0.6\textwidth]{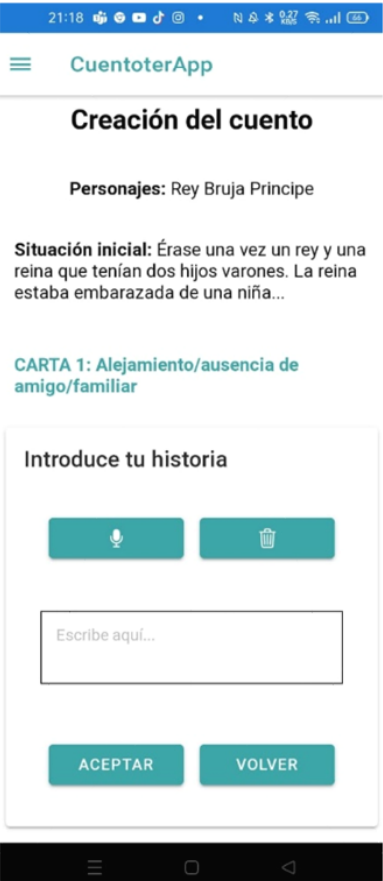}}
		\caption{Write}
		\label{fig:writeFunction}
	\end{subfigure}
    \begin{subfigure}{0.3\linewidth}
        \centering
		\fbox{\includegraphics[width=0.6\textwidth]{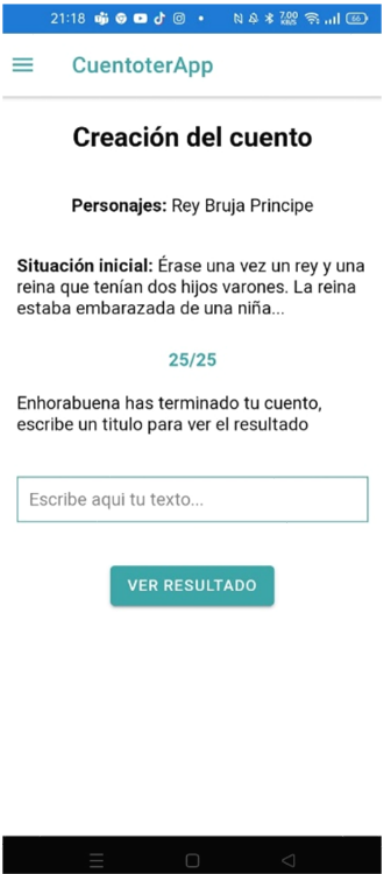}}
		\caption{Title}
		\label{fig:titlePage}
	\end{subfigure}
	\caption{Functions}
	\label{fig:appFunction}
\end{figure}

Below is the page where the story is written based on the functions (Figure \ref{fig:descriptionFunction}). Each of these functions is represented by a card with its title and a brief explanation of what it represents. Two buttons appear on the card: "WRITE" and "REJECT". If you choose to reject, the current function is simply skipped and the process moves on to the next one. However, if accepted, the card undergoes a transformation, revealing a text box for writing and a new set of buttons as shown in figure \ref{fig:writeFunction}. The ‘Microphone’ button activates voice recognition, recording everything spoken aloud into the text box. Pressing it again turns off the voice recognition. The ‘Trash’ button deletes the text box. The ‘OK’ button saves the content written for that function and proceeds to the next one. The ‘Reject’ button reverts to the view displaying the description of the function (Figure \ref{fig:descriptionFunction}).

When you have finished with all the cards, a screen appears where you are asked to enter the title for the story (see \ref{fig:titlePage}). In addition, there is a "View result" button that takes us to the next step.


\begin{figure}[H]
	\centering
    \begin{subfigure}{0.3\linewidth}
        \centering
		\fbox{\includegraphics[width=0.6\textwidth]{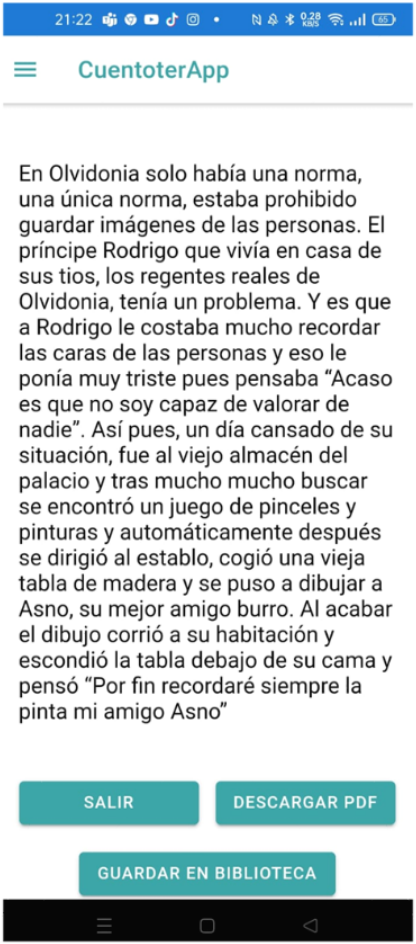}}
		\caption{Story Result}
    \label{fig:storyResult}
	\end{subfigure}
	\begin{subfigure}{0.3\linewidth}
        \centering
		\fbox{\includegraphics[width=0.6\textwidth]{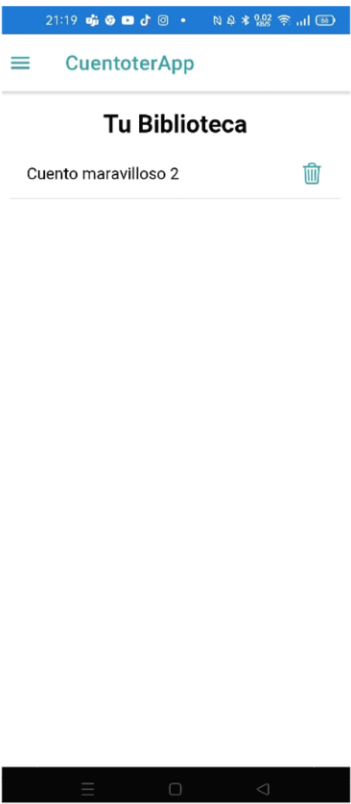}}
		\caption{Library Page}
		\label{fig:libraryPage}
	\end{subfigure}
	\begin{subfigure}{0.3\linewidth}
        \centering
		\fbox{\includegraphics[width=0.6\textwidth]{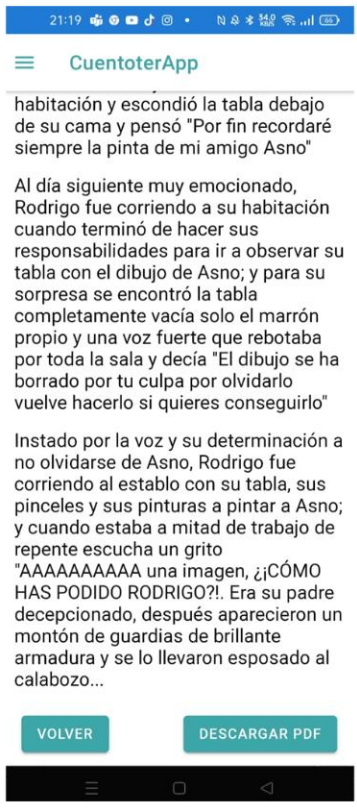}}
		\caption{Read Library}
		\label{fig:readLibrary}
	\end{subfigure}
	\caption{Library Functions}
	\label{fig:libraryFunctions}
\end{figure}

Upon clicking this button, you are directed to the result page of the story, depicted in Figure \ref{fig:storyResult}. This page allows you to read the written content and provides you with several options through a set of buttons. The ‘Exit’ button takes you back to the home screen, shown in Figure \ref{fig:welcomPage}. The ‘Download PDF’ button enables an automatic download of a pdf file containing the story’s content. Lastly, the ‘Save in Library’ button not only saves the story but also redirects you to the ‘Library’ page.

In the Library page (Figure \ref{fig:libraryPage}) we can see a list of the stories saved in the local storage, they can be deleted by clicking on the trash can button and you can click on the title to open a page (Figure \ref{fig:readLibrary}) where you can read the story and where we have the same option above to download pdf.

At any given moment, you can press the button located at the top left of the application header to reveal a drop-down menu with various options. The ‘Home’ option navigates you to the home page, as shown in figure \ref{fig:welcomPage}. The ‘Library’ option leads you to the library page, depicted in figure \ref{fig:libraryPage}. The ‘Instructions’ option directs you to the instructions page.


\section{Preliminary Usability Evaluation} \label{evaluation}

The evaluation of the system is carried out following the ISO/IEC 25000 \cite{1400-1700_isoiec_25000} standards, this groups a set of ISO standards for the evaluation and quality requirements of the software and system. Specifically, those related to performing a quality assessment in the use of the system: \begin{enumerate*}[label=\itshape\alph*\upshape)] 
    \item The quality in-use model defined in ISO/IEC 25010 \cite{iso25010} is used to obtain the characteristics and subcharacteristics to be measured.
    \item The metrics defined for the previous characteristics in ISO/IEC 25022 \cite{iso25022} are used.
    \item The Common Industrial Format (CIF) included in ISO/IEC 25062 \cite{iso25062} is used for the presentation of results.
\end{enumerate*}

The structure of this section will correspond to the different sections defined by the ISO/IEC 25062 ICF.

\subsection{Product Description and Objectives of the Evaluation}
CuentoterApp is a Progressive Web Application for assistance during the development of a Storytelling Therapy workshop. Through it the user can create personalized stories and store them.

The \textbf{objective} of this preliminary evaluation is to assess the quality of the use of the system. In this case of the complete system, since the size of the system is not large a test that goes through the whole system can be managed.

Additionally, we also have a technical objective, which is to verify that the developed system has the PWA features. For this purpose "lighthouse" is used as explained in section \ref{pwa}.

\subsection{Participants}
The evaluation was conducted with a varied group of individuals. This group included two teachers from the 'Escuela de Cuentoterapia' of Albacete, along with six other participants of varying ages and backgrounds.

This decision was made because it is taken into account that the use of teachers can provide valuable information and that taking into account that the workshop can be attended by people of a wide range of ages, it is important that the group of non-teachers is heterogeneous.

Below is a table summarizing the age, computer literacy and experience in Storytelling Therapy of the participants. Computer literacy is measured in a range of low (little knowledge), medium (some experience), and high (daily and normal use). Storytelling experience is measured in a range of none (just learned about the concept), low (heard about it before, but have not tried it as such) and high (have practical experience with it).

\begin{table}[htbp]%
  \caption{Summary of Evaluation Participants}%
  \vspace{1em}%
  \label{participantes}%
  \centering%
\resizebox{0.75\columnwidth}{!}{%
  \begin{tabular}{l l l l}%
  \hline
  Participant & Age & Computers knowledge &  Storytelling Exp.\\
  \hline
1 & 59 & medium & high \\
2 & 15 & high & none \\
3 & 55 & medium&  high\\
4 & 21 & high & none \\
5 & 53 & high & none \\
6 & 50 & low &none \\
7 & 22 & high & none \\
8 & 22 & high & none \\
 \hline
  \end{tabular}
 }
\end{table} 

\subsection{Task Description}
The tasks to be performed have been grouped in two different cases, with their corresponding subtasks. The first case is a tour through all the features of the system and the second case is the creation of a story with the particularity that in the middle the Internet connection is cut off to check if the user is aware of it and if this is a problem for the completion of the task. The different tasks in both cases are listed below.

\textbf{Case 1 Main Flow}:
\begin{enumerate*}[label=\itshape\alph*\upshape)] 
    \item Access the web and install the application, \item Start the creation of the story, \item Choose the first initial situation, \item Choose the prince and the donkey as characters, \item Reject up to the third letter and write the text specified, but using the microphone, \item Write the first letter with the specified text, \item Reject up to letter 10 and write the specified text, \item Reject up to letter 19 and write the specified text, \item Reject up to the title entry screen and type "Wonderful Story", \item View the result and save it in the library, \item Open the story in the library and download it as pdf, \item Exit from viewing the story and delete it.
\end{enumerate*}

\textbf{Case 2 Offline}: Here the device is delivered directly from the choice of the initial situation:
\begin{enumerate*}[label=\itshape\alph*\upshape)]
   \item Choose the second initial situation, \item Choose the prince and the donkey as characters,\item Reject the first letter and write the second letter with the text specified (Here in the middle of the writing process I cut the internet connection), \item Reject up to the fifth letter and write the text specified,\item Reject until the 12th letter and write the text specified, \item Reject to the title entry screen and type "Wonderful Tale 2",\item Click “salir”.
\end{enumerate*}

By way of justification, the tasks of case 1 are chosen because it is a summary of all the capabilities of the system, as well as all the possibilities of an end user and those of case 2 to highlight that feature of web application without internet and try to simulate a similar situation. The criterion for compliance is to complete all tasks, regardless of the time spent or errors made.

\subsection{Evaluation Design}
This section describes the design of the evaluation, detailing the location, the techniques and tools used, and devices, and describing the development of the evaluation.

\emph{Location}

Regarding the place of evaluation, it will be different for each participant, since it is necessary to do it in person and therefore for each one there will be different spaces.

\emph{Techniques and tools}

A stopwatch, paper and pen are simply required as tools for the evaluation.

For the correct completion of the tasks, the participants are given a sheet with the tasks to be performed and an attached sheet with the texts to be written in the corresponding part. In addition, they are given a brief explanation of the capabilities of the system, as well as of the Storytelling Therapy.

In addition, it was decided to apply the "Think Aloud" technique, which, as its name suggests, consists of expressing the thoughts we are having while performing the tasks. This way better feedback can be obtained. This is explained to the participants prior to the tasks.

\emph{Devices}

The device used will be a mobile phone *OPPO A72*, 5 GB RAM *Android* 11 version. In addition to a secondary device will be either a second mobile device that is serving as a wifi network or a router, over which there is control to cut the internet connection as desired.

\emph{Development}

The evaluation is carried out individually for each participant. Summary of Evaluation Participants and following the order of the tasks and the indications explained in the previous section.

At the beginning of the test, each person was reminded that they were undergoing a system evaluation test, not a personal test so they should not worry about the results being good or bad, since they should be as objective as possible to improve the system. They were told that they would be anonymous, they were asked to be as honest as possible and the "Think Aloud" technique was explained to them.

Afterwards, they were explained what the evaluation and the tasks consisted of, resolving any doubts clearly and concisely. And they were told that they could ask for help at any time they considered it necessary. In addition, they were asked some questions before starting the evaluation: (a) How old are you?; (b) From 1-3 with 1 being no experience and 3 being daily use experience, how would you consider your computer skills?; (c) From 1-3 being 1 not knowing it and 3 having experience with the workshop, how would you consider your knowledge of Storytelling Therapy? After the questions, the tasks are completed and the evaluation administrator is dedicated to resolving doubts and recording the different metrics. 

Finally, a SUS questionnaire is passed to them via Google Forms for the satisfaction part. The System Usability Scale (SUS) is a quick and effective tool for measuring a system’s usability from the user’s perspective. It captures users’ subjective feelings about the system’s ease of use and efficiency within the limited time typically available in an evaluation session. \cite{brooke2013sus}

\textbf{Usability metrics}

The quality model defined in ISO/IEC 25010 defines the general context of this evaluation. The metrics defined for these characteristics in ISO/IEC 25022 are used. However, from the set of characteristics found, not all of them will be taken, but the following ones: (a) Effectiveness: Rate of tasks completed without assistance and Assisted task completion rate; (b) Efficiency: Time efficiency; (c) Satisfaction:  SUS (System Usability Scale) questionnaire to measure how satisfied the user is with the system.

\subsection{Results}
The results for the two cases are shown below. Two summary tables are shown for each participant showing attendance, errors, total time, rate of tasks completed without attendance and with attendance; and time efficiency. In addition, the mean values are also grouped. Finally, the data results of the SUS questionnaire are grouped and shown.

\textbf{Case 1 Main flow}:

For this case, the target time (calculated as the time it takes an expert to perform the tasks) is 12 minutes and 45 seconds (Table \ref{resumen-caso1}).

\begin{table}[htbp]%
  \caption{Case 1 Evaluation Results}%
  \vspace{1em}%
  \label{resumen-caso1}%
  \centering%
\resizebox{\columnwidth}{!}{%
  \begin{tabular}{l l l l l l l l}%
  \hline
 & & & & &\multicolumn{2}{c}{Completeness (\%)}\\
  Participant& Time &Assist. (num) &Errors (num) &Efficiency &Without assist. &With assist. \\
  \hline
1 &28’32" & 3 & 2 &0.41 & 75 & 25 \\
2 &14’42" &1 &0 &0.80 &91.66 &8.34\\
3 &25’26" &2 &1 &0.47 &90 &10 \\
4 &13’14" &1 &0 &0.89 &90 &10 \\
5 &20’15" &1 &0 &0.58 &90 &10 \\
6 &40’18" &5 &2 &0.29 &58.33 &41.67  \\
7 &16’55" &1 &0 &0.70 &90 &10 \\
8 &13’57" &1 &0 &0.85 &90 &10 \\
\hline
Average &21’39" &2 &1 &0.62 &84 &16  \\
 \hline
  \end{tabular}
 }
\end{table}

\textbf{Case 2 Offline}: 

For this case, the target time (calculated as the time it takes an expert to perform the tasks) is 10 minutes and 5 seconds (Table \ref{resumen-caso2}).

\begin{table}[htbp]%
  \caption{Case 2 Evaluation Results}%
  \vspace{1em}%
  \label{resumen-caso2}%
  \centering%
\resizebox{\columnwidth}{!}{%
  \begin{tabular}{l l l l l l l l}%
  \hline
   & & & & &\multicolumn{2}{c}{Completeness (\%)}\\
  Participant& Time &Assist. (num) &Errors (num) &Efficiency &Without assist. &With assist. \\
  \hline
1 &13’58" &0 &0 &0.72 &100 &0 \\
2 &12’47" &0 &0 &0.79 &100 &0 \\
3 &13’06" &0 &0 &0.77 &100 &0 \\
4 &10’49" &0 &0 &0.93 &100 &0 \\
5 &15’54" &0 &0 &0.63 &100 &0 \\
6 &22’21" &0 &0 &0.52 &100 &0 \\
7 &11’51" &0 &0 &0.85 &100 &0 \\
8 &11’42" &0 &0 &0.86 &100 &0 \\

\hline
Average  &14’3" &0 &0 &0.75 &100 &0 \\
 \hline
  \end{tabular}
 }
\end{table}

The results of the participants in the SUS questionnaire are shown below. The values are shown from 0 to 100, going from lowest to highest satisfaction. The mean, minimum and maximum values are also in Table \ref{sus}.

\begin{table}[htbp]%
  \caption{SUS score}%
  \vspace{1em}%
  \label{sus}%
  \centering%
  \begin{tabular}{l l l l l l l l l}%
  \hline
  Participant & 1 & 2 & 3 &4 &5 &6& 7& 8 \\
  SUS score &97.5 &75 &82.5 &77.5 &72.5 &77.5 &82.5 &77.5\\
  \hline
\hline
Average  &80.31\\
 \hline
  \end{tabular}
\end{table}

In addition, Jeff Sauro \cite{phd_measuring}, who studied the results of more than 5000 users from 500 evaluations provides an interpretation of these scores in which he states that:
\begin{itemize}
    \item Score greater than 68 indicates the app is above average.
    \item A score less than 68, indicates below average.
\end{itemize}

\begin{figure}[H]
  \includegraphics[width=0.5\linewidth]{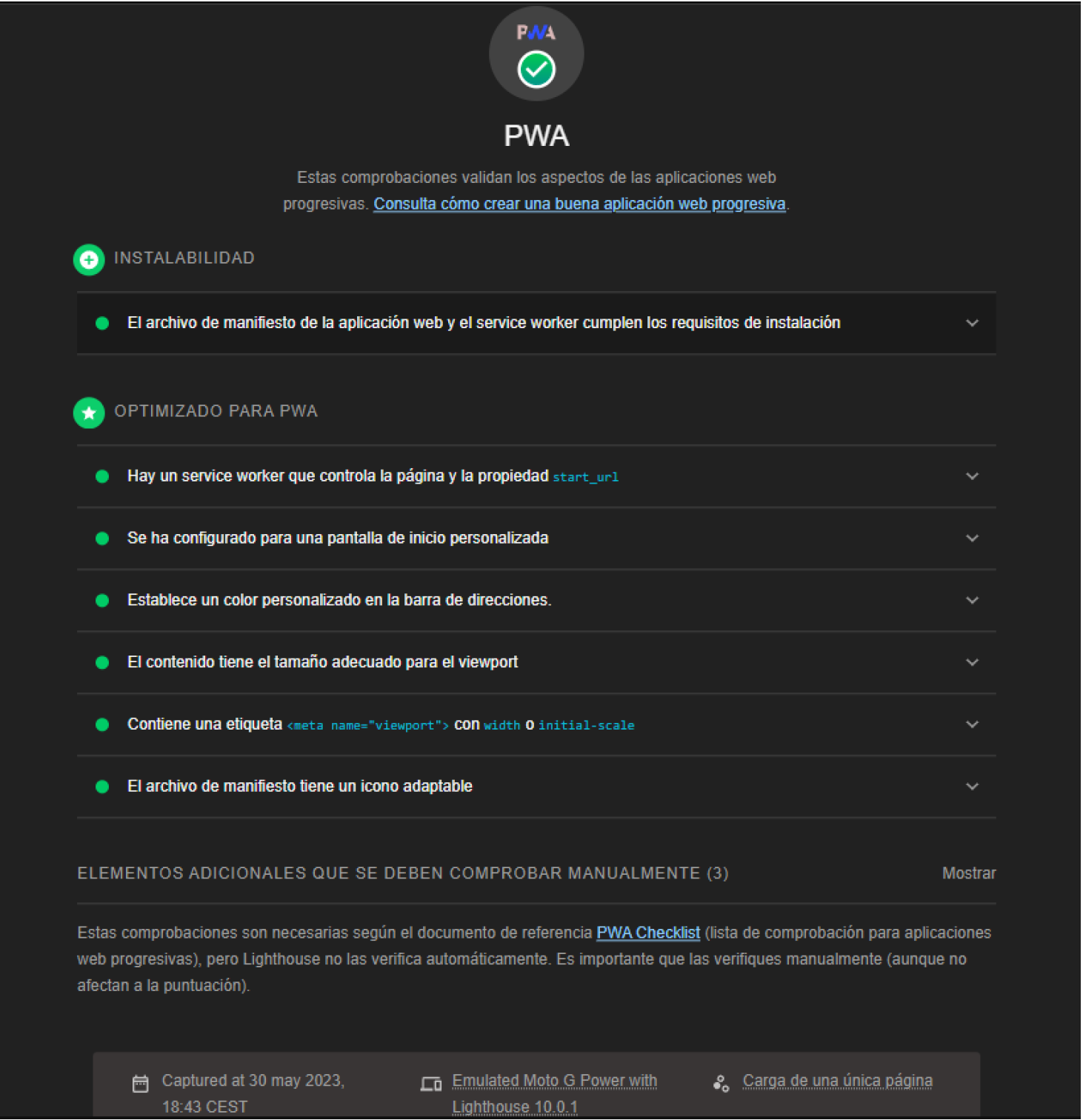}
  \caption{Lighthouse results}
  \Description{Lighthouse results}
  \label{fig:lighthouse}
\end{figure}

As for PWA, Figure \ref{fig:lighthouse} depicts a screenshot with the result of passing the PWA test provided by Lighthouse on the deployed URL.

As can be seen in the test result (see Figure \ref{fig:lighthouse}) the system passes positively in all its tests such as the ability to install itself, the registration of the service worker, the correct configuration of the Web Manifest and several style issues. So we can affirm that the developed system complies with the PWA characteristics.

Participants in the storytelling therapy association suggested that the final letter in the function letters should be mandatory, signifying every story should have an ending, preferably positive. They found handwriting on the phone cumbersome for tests, but beneficial for workshops, as it provides thinking space.

\section{Conclusions and future work} \label{conclusions}
This article presents the development of a Progressive Web Application for the support of Storytelling Therapy that we have called CuentoterApp, which allows the creation of personalized stories with assistance and their storage.

Throughout the article we have evaluated and studied the applicability of PWA to a real project, collecting which features can be really useful, which brings a lot of knowledge about progressive web applications, their advantages and disadvantages and how to implement them. This knowledge was used in developing the PWA CuentoterApp.

The article includes a validation with real end users who have provided numerous ideas for the development of future improvements. The evaluation provides relevant information on the usability of the CuentoterApp application.

Regarding the PWA technology used, an application is not a PWA in terms of "yes" or "no" but is progressive in a range, depending on the number of PWA characteristics it fulfils. These features have been defined and it has been discussed which ones were desirable.

Of those that were explained as desirable, the ability to be installed and the ability to be linkable have been met. In the case of the feature of being network independent, it is fulfilled if at least one first time the system is traversed connected to the internet, this is so due to the nature of the PWA is not limited only to the objectives of this application.

The responsive design generally adapts to all screen types, but it is primarily designed for mobile devices.

As for future work, the following improvements are noted: improving the creation of the story, working with a database as a service (cloud services); having an online library where to leave the stories; and integrating Fooocus an open-source artificial intelligence, capable of generating images with a multitude of styles from inputs, thus facilitating the generation of stories \cite{Fooocus2024}.

\begin{acks}
Special thanks to Escuela de Cuentoterapia de Albacete. This work has been partially supported by the national project granted by the Ministry of Science and Innovation (Spain) with reference PID2022-140974OB-I00 and by the regional project with reference SBPLY-21-180501-000056, granted by Junta de Comunidades de Castilla-La Mancha and the European Regional Development Funds (FEDER).
\end{acks}

\bibliographystyle{ACM-Reference-Format}
\bibliography{acmart}



\end{document}